\documentclass[a4paper,11pt]{article}
\usepackage{geometry}
\usepackage{amsmath}
\usepackage{amssymb}
\usepackage{caption}
\usepackage{comment}
\usepackage{graphicx}
\usepackage{latexsym}
\usepackage{times} 
\usepackage[english]{babel}
\usepackage{fancyhdr}
\usepackage[inkscapearea=page]{svg}
\usepackage{bm}
\usepackage[justification=centering]{caption}

\newtheorem{definition}{Definition}
\newtheorem{assumption}{Assumption}
\newtheorem{theorem}{Theorem}[section]

\addto{\captionsenglish}{\renewcommand{\abstractname}{Executive Summary}}

\fancyfootoffset{1cm}
\fancypagestyle{plain}{%
\fancyhead{}
\cfoot{}
\fancyfoot{}
\rfoot{\thepage}
\lfoot{{\small EVS38 International Electric Vehicle Symposium and Exhibition}}

}

\renewenvironment{abstract}
 {
  {\bfseries \large{\abstractname}}
  \par
  \vspace{10pt}
  \normalsize
 }

\title{\Large \emph{38th  International Electric Vehicle Symposium and Exhibition (EVS38)}\\ \emph{Göteborg, Sweden, June 15-18, 2025}\\ \hspace{10pt}\\ \LARGE\bf 
Spectral Analysis of Approximated Capacity Fade Curvature for Lithium-Ion Batteries}

\author{{\large            Huang Zhang$^{1,2}$, Torsten Wik$^2$}\\
        {\small $^1$\em Volvo Group Trucks Technology, 405 08 Gothenburg, Sweden, huang.zhang@volvo.com}\\
        {\small $^2$\em Department of Electrical Engineering, Chalmers University of Technology, 412 96 Gothenburg, Sweden}}

\date{}
\begin{document}

\setlength{\parindent}{0mm}
\baselineskip 10pt

\maketitle

\rule{\textwidth}{1pt}
\begin{abstract}
The techno-economic benefits of incorporating battery degradation into advanced control strategies necessitate the development of degradation diagnosis as an advanced function in battery management systems (BMSs). To address this, a curvature-based knee identification method was proposed in our previous work \cite{zhang2024battery}. Here, we further validate its effectiveness on a new battery aging dataset under a realistic driving profile and conduct spectral analysis of the approximated capacity fade curvature. The curvature-based method shows consistent knee identification performance on this dataset and the approximated curvature is found to correlate with underlying degradation modes and a shift of electrode material phase transition points. The method uses capacity data as the only input, which is easy to acquire in the lab and it is applicable in battery energy storage systems for grid applications.
\\

\end{abstract}

\rule{\textwidth}{1pt}
\vspace{10pt}

\section{Introduction}
Lithium-ion batteries are one of the key technologies to decarbonize transportation and power sectors due to their excellent characteristics, such as high energy density, decreasing cost, and long cycle life \cite{schmuch2018performance}. The future of battery research lies not only in the development of next-generation cell chemistries, such as solid-state \cite{janek2023challenges}, lithium-air \cite{chen2021lithium}, lithium-sulfur \cite{zhou2022formulating}, and redox-flow \cite{zhang2022emerging}, but also better management of existing lithium-ion batteries, such as lithium nickel manganese cobalt oxide (NMC), lithium manganese oxide (LMO), lithium nickel cobalt aluminum oxide (NCA), and lithium iron phosphate (LFP) \cite{schmuch2018performance}. It has been demonstrated that incorporating battery degradation into advanced control strategies can significantly improve battery lifetime and revenue in various battery applications \cite{reniers2018improving} \cite{reniers2021unlocking}. Hence, this necessitates the development of advanced functions, such as battery degradation diagnosis, in battery management systems (BMSs).
\\

To date, battery degradation diagnostic methods can generally be divided into three categories, i.e., post-mortem analysis, model-based analysis, and curve-based analysis. The post-mortem analysis is a destructive method where aged cells are disassembled in a well-controlled environment and then their degradation is examined at multiple levels through material analysis, while the other two are non-invasive diagnostic methods that identify and possibly quantify degradation mechanisms with the aid of traditional sensing techniques (e.g., voltage, current, and temperature sensors) or novel sensing techniques (e.g., acoustic, strain, and fiber-optic sensors) \cite{sommer2015monitoring} \cite{allart2018model}. Hence, non-invasive diagnostic methods that utilize physics-based models (e.g., single particle model \cite{fan2023nondestructive}) or specific measurements from reference performance tests (e.g., electrochemical impedance spectroscopy (EIS) \cite{zhang2020identifying}, pseudo open circuit voltage (OCV) \cite{birkl2017degradation}, incremental capacity analysis (ICA) \cite{dubarry2006incremental} and differential voltage analysis (DVA) \cite{bloom2005differential}) have the potential to be implemented in the BMS. However, physics-based models suffer from high computational complexity and poor identifiability, and specific measurements are challenging to acquire in the BMS. To address this, we proposed in our previous work to approximate capacity fade curvature with cyclic capacity measurements as the only input, which is much easier to acquire \cite{zhang2024battery}. As a result, the approximated curvature of battery cells with knee occurrence revealed a new oscillatory degradation phenomenon over a period, which separated the degradation process into three distinct phases. The oscillatory degradation phenomenon was then used to identify the knee and its onset for those cells. This work is an extension of our previous work with new contributions as follows:
\begin{itemize}
\item 
We further validate the effectiveness of our previously proposed knee identification method on a new experimental battery aging dataset under a realistic driving profile and correlate the identified knee and its onset with estimated battery degradation modes. 
\item 
We conduct spectral analysis of approximated capacity fade curvature in three degradation phases and find out that the spectral of approximated curvature in phase 2 exhibits steady fluctuation over sampled frequencies without dominating peaks. Therefore, we associate the significant fluctuation of approximated curvature in phase 2 with electrode material phase changes and find a strongly correlated shift of phase transition points using incremental capacity curves.
\end{itemize}

\section{Data Curation}
\begin{figure}[htbp!]
  \begin{centering}
  \includegraphics[width=72mm]{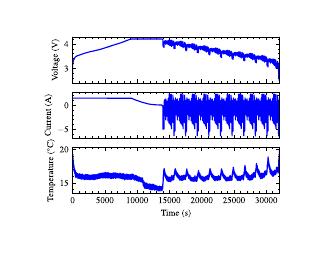}
  \caption{One full charge-discharge cycle of a sample cell in experiment 4 of ICL dataset.}
  \label{fig1}
  \end{centering}
\end{figure}

The battery aging dataset was generated by Imperial College London (ICL) \cite{kirkaldy2024lithium}, which consists of 40 NMC 811/graphite-$\text{SiO}_\text{x}$ cylindrical cells from LG Chem (model GBM50T2170, 5 Ah nominal capacity). The dataset is divided into 5 experiments according to different state-of-charge (SoC) windows (or discharge profiles). Each experiment contains data of cells aged at 3 different ambient temperatures (10$^{\circ}$C, 25$^{\circ}$C, and 40$^{\circ}$C).
In particular, experiment 4 has 8 cells that were charged with a uniform 0.3C CC-CV charging step and then discharged with the World wide harmonized Light vehicle Test Protocol (WLTP) driving profile from 100\% SoC to 0\% SoC (see Fig. \ref{fig1}).

\begin{figure}[htbp!]
  \begin{centering}
  \includegraphics[width=72mm]{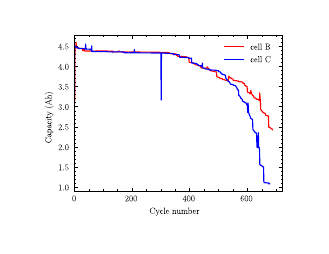}
  \caption{Cyclic capacity data of cell B and C in experiment 4 of ICL dataset.}
  \label{fig3}
  \end{centering}
\end{figure}

In terms of battery capacity measurements, both cyclic capacity (0.3C charge, 10$^{\circ}$C) from aging tests and capacity (C/10 discharge, 25$^{\circ}$C) from reference performance tests (RPTs) are available. Two cells in experiment 4 have knee occurrence on their capacity fade curves. This is likely due to lithium plating at the negative electrode as these two cells were cycled at low temperature (10$^{\circ}$C) \cite{kirkaldy2024lithium}. As illustrated in Fig. \ref{fig3}, the cyclic capacity data of two cells, i.e., cell B and C, will be used to identify the knee and its onset.

\newpage
\section{Methods} 
This section will introduce our previously proposed capacity knee identification method and the power spectral density estimation method used in this work.
\subsection{Curvature-based capacity knee identification method}
Given a sequence of capacity states of a lithium-ion battery, we can regard it as a discrete-time stochastic process denoted by $\bm{x}(n), \ n=0, 1, \ldots, N-1$. Hence a time series capacity measured per cycle can be viewed as a sample or realization of the given process with a sampling interval $\Delta t = 1 \ \mathrm{cycle}$ (see Fig. \ref{fig3}). Hence the sampling frequency $f_s = 1/\Delta t$ in $\text{cycle}^{-1}$ or $\omega_s=2\pi/\Delta t$ in $\text{rad}\cdot \text{cycle}^{-1}$. Note that we will use the notation $\bm{x}(n)$ to represent both a discrete-time stochastic process and a realization of the given process (i.e., a time series). 
\\

To identify knee and knee-onset points on the battery capacity fade curve, a curvature-based identification method was proposed in our previous work \cite{zhang2024battery}. The concept of curvature is a mathematical measure of the amount by which a curve deviates from being a straight line \cite{satopaa2011finding}. Hence, a knee on the capacity fade curve can be defined as the point of maximum curvature. In practice, the curvature must first be approximated from time series capacity data. To begin with, two assumptions are required by our curvature-based identification method:
\begin{assumption}
    The lithium-ion battery has a knee occurred on its capacity fade curve.
\end{assumption}
\begin{assumption}
    The capacity data is evenly sampled. If not, the data points are fitted to a spline function and interpolated to become so.
\end{assumption}
Then we summarize this method in a step-by-step manner as follows:
\begin{enumerate}
\item 
Normalize the time series capacity data with the battery's initial nominal capacity.
\item 
Smooth the normalized time series capacity data using the Savitzky-Golay filter.
\item 
Calculate the discrete capacity fade curvature at each data point of the smoothed data.
\item 
Identify knee and knee-onset points using a time series segmentation algorithm, for example, Fast Low-cost Unipotent Semantic Segmentation (FLUSS) algorithm \cite{gharghabi2017matrix}.
\end{enumerate} 
The identified knee and knee-onset points divide the battery degradation process into three discrete phases
\begin{itemize}
\item[] 
\textbf{Phase 1}: From the beginning of life to the knee-onset point.
\item[] 
\textbf{Phase 2}: From the knee-onset point to the knee point.
\item[] 
\textbf{Phase 3}: From the knee point to the end of life.
\end{itemize}

\subsection{Power spectral density estimation method}
For a time series, one can calculate the power spectral density, which describes how the power of a time-series signal is distributed over frequencies. Here we are only concerned with stochastic processes that are wide sense stationary (WSS) and ergodic, which means their mean values are constant, their auto-correlations only depend on the time lag, and their time averages converge to ensemble averages \cite{papoulis2002probability}. Then the power spectral density (PSD) of a WSS process $\bm{x}(n)$ is the discrete-time Fourier transform (DTFT) of its auto-correlation.
\\

Determining the exact DTFT requires an infinite number of samples while in practice only a finite number of samples are available for analysis. Therefore, we will work with a time series with indices $n=0, 1, \ldots, N-1$. Mathematically, this can be described as
\begin{equation}\label{win_DTFT}
    \hat{\bm{x}}(n) = \bm{x}(n) r(n), \ \forall n   
\end{equation}
where $r(n)$ is the window function. For example, a rectangular window function is defined as
\begin{equation}\label{rec_win}
    r(n) = 
    \begin{cases}
    1, \ \mathrm{if} \ n =  0, 1, \ldots, N-1 \\
    0, \ \mathrm{otherwise}.
    \end{cases}
\end{equation}
The rectangularly windowed DTFT can then be derived from Eqn. (\ref{win_DTFT}) and given as
\begin{equation}
    \hat{\bm{X}}(\omega) = \frac{1}{\omega_s}\int^{\omega_s}_{0} \bm{X}(\lambda) R(\omega-\lambda) d\lambda.
\end{equation}
If we sample the windowed DTFT at evenly-spaced frequencies $\omega_k \Delta t = 2\pi k/N$, we will obtain the windowed discrete Fourier transform (DFT)
\begin{equation}\label{win_DFT}
    \bm{X}(k) = \hat{\bm{X}}(\frac{2\pi k}{N \Delta t}) = \sum^{N-1}_{n=0} \bm{x}(n) r(n)e^{-j 2\pi n k/N},
\end{equation}
where $k = 0, 1, \ldots, N-1$ is the frequency point. 
Following Parseval's theorem, the energy of the time series $\bm{x}(n)$ can be calculated from its windowed DFT $\bm{X}(k)$ by
\begin{equation}
    E_{\bm{X}} = \Delta t \sum^{N-1}_{k=0} \Big| \frac{\bm{X}(k)}{C} \Big|^2,
\end{equation}
where $C$ is a normalization constant defined as
\begin{equation}
    C=\sqrt{\sum_{n=0}^{N-1}|r(n)|^2}.
\end{equation}
The power of the time series can be calculated by
\begin{equation}
    P_{\bm{X}} = \frac{1}{N \Delta t} E_{\bm{X}} = \frac{1}{N} \sum^{N-1}_{k=0} \Big| \frac{\bm{X}(k)}{C} \Big|^2.
\end{equation}
Lastly, the windowed frequency-discrete PSD of the time series is defined as
\begin{equation}
    S(k) = \Delta t \Big| \frac{\bm{X}(k)}{C} \Big|^2.
\end{equation}

In practice, the PSD of a stochastic process can be estimated using non-parametric methods, such as the periodogram \cite{schuster1898investigation}, the modified periodogram \cite{papoulis2002probability}, Bartlett's method \cite{bartlett1948smoothing}, Blackman-Tukey \cite{stoica2005spectral}, and Welch's method \cite{welch1967use}. Note that all the non-parametric methods are modifications of the classical periodogram \cite{schuster1898investigation}. The periodogram is defined as \cite{proakis2007digital}
\begin{equation}
    S_N(k) = \frac{1}{N} \Big| \sum^{N-1}_{n=0} \bm{x}(n) e^{-j 2\pi n k/N} \Big|^2.
\end{equation}
It can be seen from the equation above that the periodogram uses a rectangular window function defined in Eqn. (\ref{rec_win}) to reduce the leakage effects in the estimation. The modified periodogram uses a non-rectangular window function. However, it is well-known that the variance of the periodogram does not tend to zero as the number of samples tends to infinity. Hence the periodogram is an inconsistent estimator of the PSD. A way to enforce the decreasing variance is called averaging. For example, Bartlett's method first divides the time series of length $N$ into $K$ segments of length $M$ each, then computes the periodogram for each segment, and lastly averages the result of the periodograms for the $K$ segments. Compared with the classical periodogram, Bartlett's method reduces the variance of the periodogram at the expense of a reduction of spectral resolution \cite{bartlett1950periodogram}. Welch's method resolves the trade-off between variance and spectral resolution in Barlett's method by applying a window function to each overlapping segment \cite{welch1967use}. Considering many successful applications of Welch's method reported in the literature, Welch's method is chosen for the estimate of PSD in this work.

\section{Results and Discussion}
The effectiveness of our proposed identification method has been validated on two experimental battery aging datasets of two different chemistries (LFP/graphite and NMC 811/graphite) in two different operating conditions and one synthetic dataset of NMC 811/graphite-$\text{SiO}_\text{x}$ batteries in one operating condition in our previous work \cite{zhang2024battery}. In this section, we will first validate our proposed identification method on one additional experimental battery aging dataset and correlate the identified knee and its onset with battery internal degradation modes and the electrode material phase transitions during charge and discharge. Then, the spectral analysis of approximated capacity fade curvature in three degradation phases will be provided for these cells with knee occurrence.

\subsection{Knee and knee-onset identification performance}
\begin{figure}[htbp!]
  \begin{centering}
  \includegraphics[width=72mm]{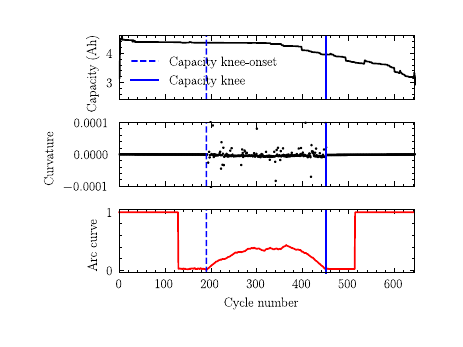}
  \includegraphics[width=72mm]{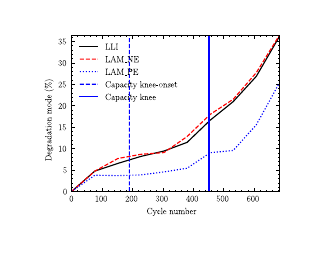}
  \caption{The approximated curvature (left) and internal degradation modes (right) for cell B. The knee-onset and knee points are identified at the 190th and 452rd cycle, respectively.}
  \label{fig4}
  \end{centering}
\end{figure}

\begin{figure}[htbp!]
  \begin{centering}
  \includegraphics[width=72mm]{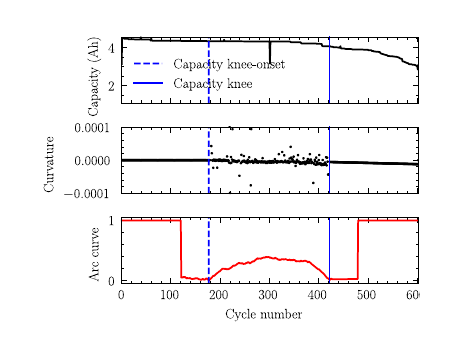}
  \includegraphics[width=72mm]{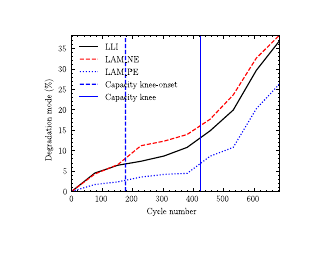}
  \caption{The approximated curvature (left) and internal degradation modes (right) for cell C. The knee-onset and knee points are identified at the 177th and 423rd cycle, respectively.}
  \label{fig5}
  \end{centering}
\end{figure}

As a measure of the rate of change of capacity fade rate, the approximated curvature with its corresponding arc curve is plotted for two cells with knee occurrence in experiment 4 of ICL dataset, i.e., cell B in Fig. \ref{fig4} and cell C in Fig. \ref{fig5}, respectively. In addition, three battery internal degradation modes, i.e., loss of lithium inventory (LLI), loss of active material at the negative electrode (LAM\_NE), and loss of active material at the positive electrode (LAM\_PE), were estimated by Kirkaldy et al. \cite{kirkaldy2024lithium}, using pseudo-open circuit voltage (pseudo-OCV).
It can be seen in both Fig. \ref{fig4} and Fig. \ref{fig5} that the approximated curvature fluctuates significantly in phase 2 of which the boundaries [blue lines] were inferred by the proposed identification method. Moreover, the height of the arc curve is the lowest at the two boundaries. The arc curve specifies the number of nearest neighbor arcs that cross over each location \cite{gharghabi2017matrix}. Since most of the approximate curvature subsequences shall find their nearest neighbors within the same phase, we would expect very few arcs to cross over the boundaries [blue lines]. That rationalizes why the height of the arc curve is the lowest at the two boundaries. In addition, all three degradation modes contribute to the loss of capacity, i.e., they all begin with a square root dependence on time until they reach the inflection point identified as the knee-onset point, and then accelerate exponentially, which leads to knee occurrence on their capacity fade curves.

\begin{figure}[htbp!]
  \begin{centering}
  \includegraphics[width=72mm]{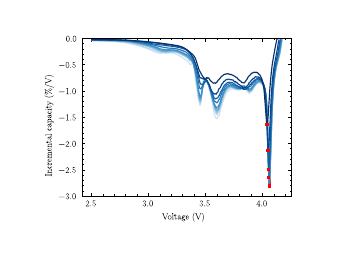}
  \includegraphics[width=72mm]{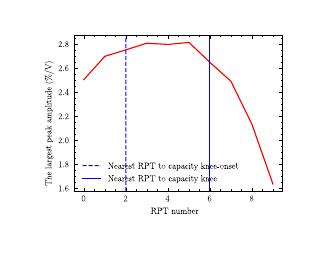}
  \caption{The IC curves with largest peaks marked with solid red squares (left) and amplitudes of largest peaks (right) from 10 RPTs for cell B.}
  \label{fig_8}
  \end{centering}
\end{figure}

\begin{figure}[htbp!]
  \begin{centering}
  \includegraphics[width=72mm]{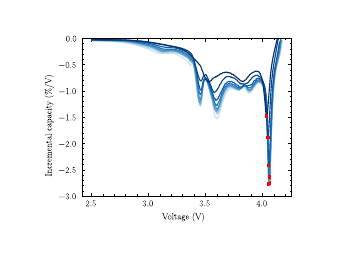}
  \includegraphics[width=72mm]{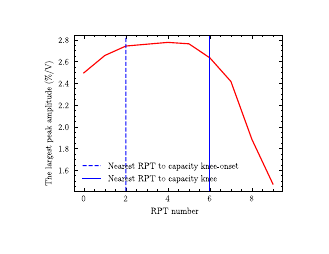}
  \caption{The IC curves with largest peaks marked with solid red squares (left) and amplitudes of largest peaks (right) from 10 RPTs for cell C.}
  \label{fig_9}
  \end{centering}
\end{figure}

\newpage
We hypothesize that the fluctuation of approximated curvature in phase 2 (i.e., from the knee-onset point to the knee point) may be associated with electrode phase changes (also known as intercalation stage transitions) over the course of charge and discharge \cite{sommer2015monitoring} \cite{allart2018model}.
The functional principle of lithium-ion batteries is built upon lithium intercalation and deintercalation, which leads to structural changes of the electrode active materials and consequently cell volume expansion and contraction. For most lithium-ion chemistries, the anode material dominates the cell volume change, such that the cell expands during charge and contracts during discharge. With lithium intercalated into the anode active material during charge, different crystallographic phases are formed through several stages until the negative electrode is fully charged \cite{allart2018model}. 
In Ref. \cite{sommer2015monitoring}, it was reported that the voltage signatures at which phase transitions occur change with cell aging. The phase transitions in the electrode active material can be identified from the detected peaks in the derivative of the cell OCV. Thus, the incremental capacity (IC) curve is used here as its peaks correspond to the phase transitions \cite{barai2019comparison}. It can be seen in Fig. \ref{fig_8} and Fig. \ref{fig_9} that the amplitudes of the largest peak at around 4.05V first increase until they reach a saturation point close to the capacity knee-onset point and then start to decrease around the capacity knee point. The observed changes of the peak amplitude have been reported to be strongly associated with the cell’s homogeneity of lithium distribution \cite{xie2022inhomogeneous}. The inhomogeneous distribution of current density in local units will induce asynchronous intercalations in the electrode active material, which misplaces the IC peak of each local unit. The superposition of misplaced peaks results in the decrease of total peak amplitude. However, the changes of the peak amplitude in IC curves may not only be caused by the homogeneity in the electrode, but also by other degradation mechanisms. 

\newpage
We further validated the effectiveness of the curvature-based knee identification method proposed in our previous work on an experimental battery aging dataset and correlated the identified knee and its onset with estimated battery internal degradation modes and the electrode phase transitions during charge and discharge. So far the effectiveness of our proposed capacity knee identification method has been validated on four battery aging datasets of three different chemistries (i.e., LFP/graphite, NMC 811/graphite, and NMC 811/graphite-$\text{SiO}_\text{x}$) in three different operating conditions (i.e., fast charging protocols/galvanostatic discharge profile at 30$^{\circ}$C, CC-CV charging protocols/galvanostatic discharge profile at 15$^{\circ}$C, 25$^{\circ}$C and 35$^{\circ}$C, and CC-CV charging protocols/WLTP discharge profile at 10$^{\circ}$C) \cite{zhang2024battery}.
The advantages of this curvature-based knee identification method over existing identification methods in the literature are two-fold: 1) it applies to battery capacity fade curves with a wide range of shapes, which can be a combination of linear, sublinear, and superlinear curves \cite{attia2022knees}; 2) it uses capacity fade data as the only input, which is data efficient and thus preferable in practical applications, such as classification of retired electric vehicle batteries for second-life applications.

\subsection{Curvature spectral analysis}
\begin{figure}[htbp!]
  \begin{centering}
  \includegraphics[width=72mm]{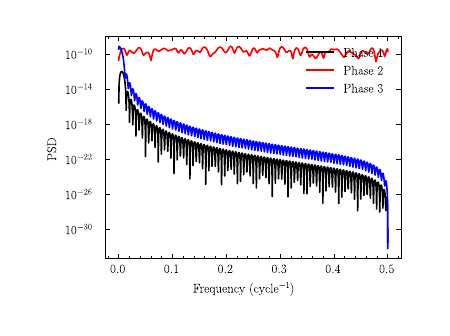}
  \includegraphics[width=72mm]{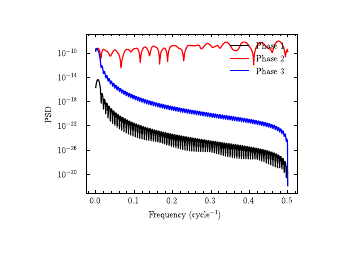}
  \caption{PSD of approximated curvature in three phases for cell B (left) and cell C (right) in experiment 4.}
  \label{fig6}
  \end{centering}
\end{figure}

Welch's method is considered as the modified periodogram which first applies a window function to each overlapping segment and then averages the results of the periodograms for all segments as the estimate of PSD \cite{welch1967use}. In principle, different window functions will yield different trade-offs between the main lobe width and the side lobe level. Therefore, to reduce the leakage effect without losing much frequency resolution, a Hann window function is applied for the averaging with an overlapping of 50\%. 
The resulting PSD of approximated curvature in the three phases of the battery degradation process is illustrated in Fig. \ref{fig6}. 
It can be seen that the power of the approximated curvature in phase 2 is several orders of magnitude higher than that in phases 1 and 3 over all the frequencies for both cells. Interestingly, the spectral density of the approximated curvature in phase 2 is also relatively stable, around $10^{-10}$, over the frequencies without any dominating peaks, which implies that the approximate curvature in phase 2 comprises a complex mixture of components at multiple frequencies rather than a single periodic component. In contrast, the spectral density of the approximate curvature in phases 1 and 3 exhibits an overall trend of decreasing power from low to high frequencies, with a peak at around 0.005 cycle$^{-1}$.

\newpage
\section{Conclusions}
It has been shown in the literature that incorporating battery degradation into advanced control strategies can significantly improve battery lifetime and revenue in various battery applications. This necessitates the development of battery degradation diagnosis as an advanced function in a battery management system (BMS). 
To address this, we have proposed approximate capacity fade curvature in our previous work with cyclic capacity measurements as the only input to identify the knee and its onset points. In this work, we further validate the effectiveness of this method on an additional experimental battery aging dataset in a realistic driving profile, correlated with estimated battery degradation modes. A shift of electrode material phase transition points was found to strongly correlate with the significant fluctuation of approximated curvature in phase 2. Moreover, the spectrum of the approximated curvature in phase 2 exhibited steady fluctuation over sampled frequencies without dominating peaks. 
\\

These findings lead to future work, to further validate the effectiveness of this curvature-based knee identification method on state-of-capacity (SoQ) data generated by the BMS in the field. To quantify the effects of various battery degradation parameters on the approximated curvature, parametric analysis using physics-based models should be explored.

\section*{Acknowledgments} 
This work was supported by the Swedish Energy Agency (Grant number P2024-00998). In particular, the authors would like to thank Eibar Flores from SINTEF Industry for the constructive discussion.


\bibliographystyle{evs.bst}   
\bibliography{References}



\bigskip
\section*{Presenter Biography}
\begin{minipage}[b]{21mm}
\includegraphics[width=20mm]{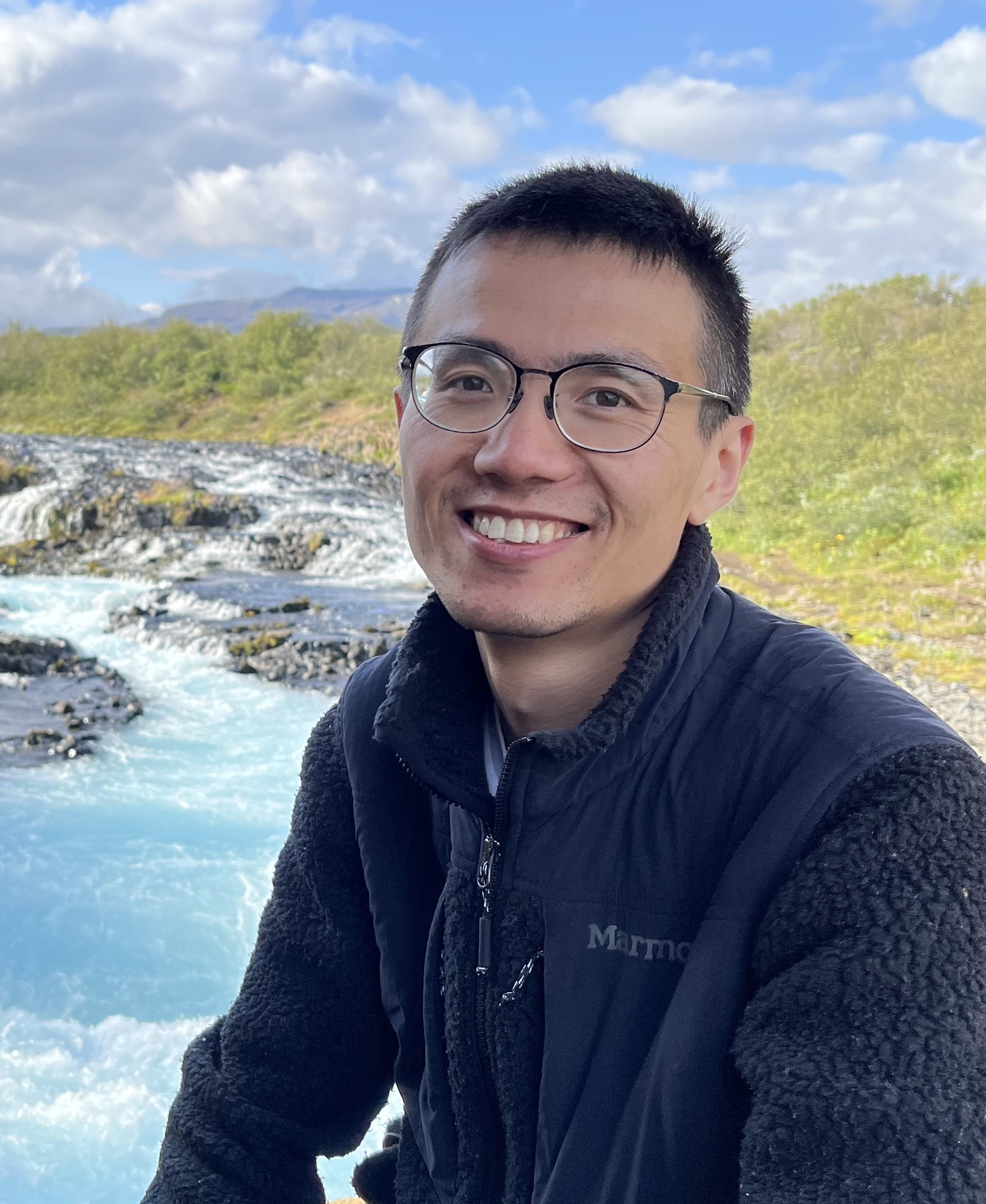}
\end{minipage}
\hfill
\begin{minipage}[b]{140mm}
\small
Huang Zhang received his M.Sc. degree in Electrical Engineering from KTH Royal Institute of Technology, and M.Sc. degree in Energy Engineering and Management from Instituto Superior Técnico in 2017. He is currently a PhD student at the Department of Electrical Engineering, Chalmers University of Technology. His research interests include uncertainty estimation in machine learning and optimal control. His current research applications include battery lifetime prediction, battery degradation diagnosis, and techno-economic modeling.
\end{minipage}
\\\\

\end{document}